\documentclass[journal]{IEEEtran}
\ifCLASSINFOpdf
\else
\fi

\usepackage{cite}

\usepackage{times}  
\usepackage{helvet} 
\usepackage{courier}  
\usepackage{amsfonts}
\usepackage{amsmath}
\usepackage{booktabs}
\usepackage{subfig}
\usepackage[hyphens]{url}  
\usepackage{graphicx} 
\usepackage{graphicx}  
\title{Decoding Spiking Mechanism with Dynamic Learning on Neuron Population}
\begin{document}
\author{
\IEEEauthorblockN{Zhijie Chen, Junchi Yan\IEEEauthorrefmark{1}, Longyuan Li and Xiaokang Yang~\IEEEmembership{Fellow,~IEEE}} 
  
\IEEEauthorblockA{MoE Key Lab of Artificial Intelligence, Shanghai Jiao Tong University} 

\IEEEauthorblockA{\{chen-zhijie, yanjunchi, jeffli, xkyang\}@sjtu.edu.cn} 
}

\maketitle

\begin{abstract}
A main concern in cognitive neuroscience is to decode the overt neural spike train observations and infer latent representations under neural circuits. However, traditional methods entail strong prior on network structure and hardly meet the demand for real spike data. Here we propose a novel neural network approach called Neuron Activation Network that extracts neural information explicitly from single trial neuron population spike trains. Our proposed method consists of a spatiotemporal learning procedure on sensory environment and a message passing mechanism on population graph, followed by a neuron activation process in a recursive fashion. Our model is aimed to reconstruct neuron information while inferring representations of neuron spiking states.  We apply our model to retinal ganglion cells and the experimental results suggest that our model holds a more potent capability in generating neural spike sequences with high fidelity than the state-of-the-art methods, as well as being more expressive and having potential to disclose latent spiking mechanism. The source code will be released with the final paper.
\end{abstract}


\section{Introduction and Related Work}
\label{sec:intro}
Spiking activity of a neuron population contains a large amount of cerebral information~\cite{salinas2001correlated,bialek1992reliability} and has been studied extensively in neuroscience. Decoding the relationship between neural spike trains and latent structure in neuron circuits can advance therapeutic neuro-technologies such as closed-loop deep brain stimulation~\cite{rosin2011closed} and brain-machine interface (BMI)~\cite{sakellaridi2019intrinsic,serruya2002brain,pandarinath2018inferring}. As a symbolization of intricate designs in human physiology, the spike train of a neuron is dependent upon multiple factors like sensory environment and the latent structure of neuron population~\cite{fourcaud2003spike,fremaux2016neuromodulated}. How these factors are combined and ultimately act on neural spikes is still not very clear and under heated discussion. 

Existing methods can be categorized into two groups. Supervised learning techniques are used to study neuron spikes. An enlightening work is the linear-nonlinear-Poisson (LNP) cascade model~\cite{truccolo2005point,paninski2004maximum}, in which spikes are generated according to an inhomogeneous Poisson process, with rate determined by an instantaneous nonlinear function of the filtered input. Based on LNP, another well-known model, the generalized linear model (GLM)~\cite{pillow2008spatio} is proposed. GLM consists of a set of linear filters followed by a nonlinearity and probabilistic spike generation. The components of GLM strictly conforms to certain phenomenon in neural spikes, which offers a concise description of neural population activity. Among all the derivatives and optimization of GLM~\cite{truccolo2005point,vidne2012modeling,weber2017capturing,zoltowski2018scaling}, pyGLM~\cite{linderman2016bayesian} combines latent variable network models with GLM in a hierarchical Bayesian framework. However, all the above model-based approaches have limitations as they entail strong prior on neuron population structure and hence are susceptible to the selection of hyperparameters. 

Other endeavors involve unsupervised learning on neuron spike data. The common route is to discover nonlinear mapping from latent variable to neural responses, where the low-dimensional variable evolves in time according to a Gaussian process prior ~\cite{wu2017gaussian,gao2015high,wu2018learning,duncker2018temporal}. These models are efficient in extracting latent variables from population spikes in relatively small sizes, but exhibit less satisfying efficacy in dealing with large scale data, which will be the common case in upcoming revolution in large-scale neuron data collection. It reminds us of a fact that an inductive model with stronger scalability is needed in practical use.

Recently, deep neural networks have been used in cognitive neuroscience and primarily validated its feasibility in parametrizing nonlinear mapping from latent space to spike rates. One basic idea is fitting a recurrent neural network for sequential analysis~\cite{batty2016multilayer,pitti2017inferno}. A similar way is to use convolutional neural networks to find relationship between stimulus and neuron response~\cite{mcintosh2016deep}. Nevertheless, these models focus on individual neuron activities and need fitting separately for different cell types. These black-box models also have difficulty in extracting underlying representations w.r.t. contributing factors. To deal with the defects in current literature, this paper is motivated to adopt a novel neural network approach, thereby separately modeling the effects of sensory environment, spiking history and surrounding neuron activities. This paper has the following contributions:

1) We propose neuron activation process where hidden state determines spiking activity of a single neuron. We combine log likelihoood of spike sequences and mutual information to capture local and global patterns in spike generation.

2) We devise a deep network approach to extract neural information as latent variables under neural circuits. Our method involves a spatiotemporal learning of the sensory environment, together with an online neuron relation learning process that supports dynamic neuron population graphs.

3) Extensive experimental results show that our proposed method has a potent capability in generating simulation neural spikes with high fidelity and is instructive in inferring representations of neuron hidden state. Our proposed method also takes the first step in explicitly disclosing latent neuron structure and spike eliciting mechanism.

\begin{figure*}[tb!]
  \centering
  \includegraphics[width=1.0\textwidth]{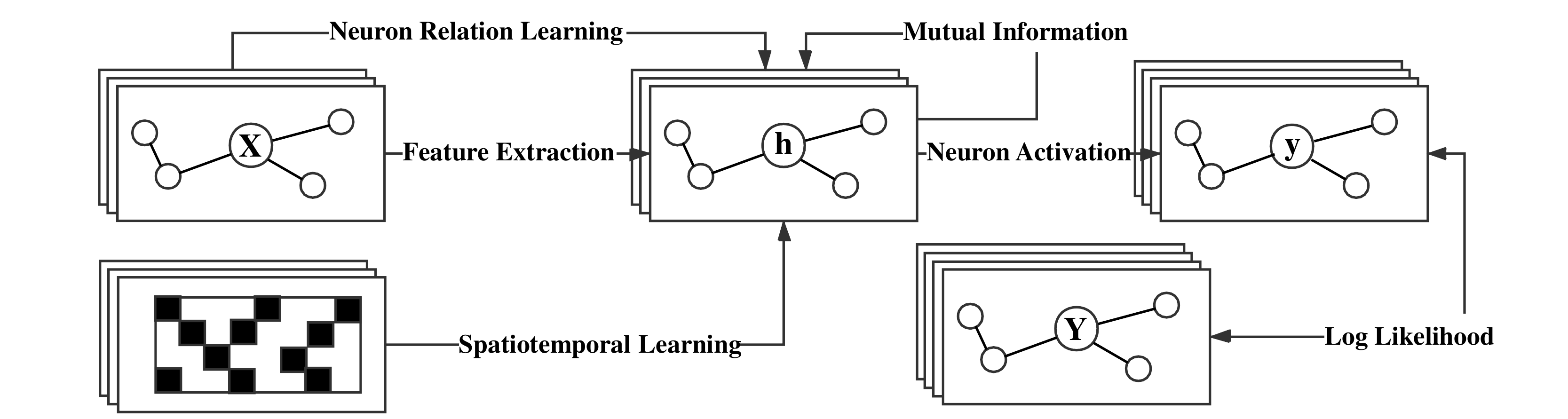}
   \vspace{-15pt}
  \caption{Neuron Activation Network (NAN). It consists of neuron relation and spatiotemporal learning to extract hidden states from the input spike sequences. The neuron hidden state is mapped to spike firing rate with neuron activation process.}
  \label{fig:workflow}
\end{figure*}

\section{Proposed Approach}
\subsection{Notations and Problem Formulation}
Suppose we have recorded the spike trains from $N$ neurons. Let $\mathbf{X} \in \mathbb{R}^{N \times T}$ denote the matrix of spike count data, with neurons indexed by $i \in \{1,2,..., N\}$ and spike counted in discrete time bins indexed by $t \in \{1,2,...T\}$. The stimulus that elicits the spike trains is denoted as $\mathbf{\Psi} \in \mathbb{R}^{W \times H \times T}$, which is a three-dimensional matrix, with the first two dimensions standing for spatial width and height respectively, and the third dimension stands for recording time bins in accord with neuron spikes. Our goal is to construct a model that grasps the latent mechanism in spike train generation and make forward prediction for a spike sequence.

Assuming prediction step to be $\tau$, i.e. the length of prediction spike sequence, $\mathbf{Y} \in \mathbb{R}^{N \times \tau}$ is the prediction matrix of spike sequence in the next $\tau$ time bins given $\mathbf{X}$. $\mathbf{E} \in \mathbb{R}^{N\times N}$ denotes the learned relation intensity matrix of neuron couples, and $\mathbf{S} \in \mathbb{R}^{N \times \tau \times n_p}$ is the stimulus factor matrix, elements of which are $n_p$-dimension embeddings of input stimulus. Being consistent with former researches, we assume the spike activity of a single neuron is determined by its hidden state variable $\mathbf{h}_{i,t} \in \mathbb{R}^{n_h}$, which is an $n_h$-element vector. $\mathbf{H} \in \mathbb{R}^{N \times T \times n_h}$ denotes the hidden state matrix of neuron population throughout the whole recording time.

With the settings stated above, we can now derive the joint distribution over prediction spike sequence and other latent variables or learned representations:
\begin{align}\label{mle}
&p(\mathbf{Y},\mathbf{H},\mathbf{E},\mathbf{S}|\{ \mathbf{\Psi}, \mathbf{X}\})\\\notag =&p(\mathbf{Y}|\mathbf{H})p(\mathbf{H}|\mathbf{E},\mathbf{S},\mathbf{X})p(\mathbf{E}|\mathbf{X})p(\mathbf{S}|\mathbf{\Psi})\\\notag
=&\prod_{i=1}^N\prod_{t=1}^\tau p(y_{i,t}|h_{i,t})p(h_{i,t}|\mathbf{E},s_{i,t},h_{i,t-1},...,h_{i,1},\mathbf{X})\\\notag
&\prod_{i=1}^N\prod_{j=1}^Np(e_{ij}|\mathbf{X}_i,\mathbf{X}_j)\prod_{i=1}^N\prod_{t=1}^\tau p(s_{i,t}|\mathbf{\Psi}_{:,:,t})
\end{align}
where $\mathbf{X}_i$ and $\mathbf{X}_j$ represent the $i$-th and $j$-th row of matrix $\mathbf{X}$, namely the neuron spike observations of neuron $i$ and $j$. Our goal is to maximize the log marginal likelihood associated with the joint distribution in Eq.~\ref{mle} as well as to get a convincing estimate for $\mathbf{H}$, $\mathbf{E}$, $\mathbf{S}$, respectively. To this end, all latent variables and representations are required to be marginalized out:
\begin{align}\label{loglikelihood}
\log~ p(\mathbf{Y})=&\log \iiint p(\mathbf{Y},\mathbf{H},\mathbf{E},\mathbf{S})d\mathbf{H}d\mathbf{E}d\mathbf{S}\\\notag
= &\log \int  p(\mathbf{Y}|\mathbf{H})\iint p(\mathbf{H}|\mathbf{E},\mathbf{S})\int p(\mathbf{E}|\mathbf{X})p(\mathbf{X})\\\notag
&\quad \int p(\mathbf{S}|\mathbf{\Psi})p(\mathbf{\Psi})d\mathbf{\Psi}d\mathbf{X}d\mathbf{S}d\mathbf{E}d\mathbf{H}
\end{align}

As the latent variables in a neuron population largely interplay with each other, using the traditional coordinate descent for learning can be inefficient and may lead to local optima~\cite{wu2017gaussian}. Therefore, in the following parts, we are going to separately model the terms in Eq.~\ref{loglikelihood} with a novel neural network approach.

\subsection{Neuron Activation Process}
 Let $\lambda_{i,t} = f(\mathbf{h}_{i,t})$ denote the firing rate of neuron $i$ at time $t$. Here neuron firing rate is in unit of spikes per time bin, and $f$ is a shared internal function mapping the $n_h$-dimensional representation of hidden state $\mathbf{h}_{i,t}$ to a scalar firing rate. The spike count of neuron $i$ at $t$ follows a Poisson distribution with its mean equal to $\lambda_{i,t}$:
$$y_{i,t}|f,\mathbf{h}_{i,t} \sim {\rm Poisson}(f(\mathbf{h}_{i,t}))$$

The hidden state $\mathbf{h}_{i,t}$ is updated for each interval to capture the effect of time-varying factors:

\textbf{Sensory environment.} Assume the neurons are continuously exposed to a movie during recording. The movie works as an exogenous stimulus for spike train generation;

\textbf{Spike history.} Spike trains are strongly time-dependent, one of the evidences is the existence of refractory period. In our implementation, a neuron's firing rate evolves with respect to the previous hidden state in a recursive fashion;

\textbf{Surrounding neuron states.} As an important way for message passing in neuron population, the hidden state of a neuron not only determines its own firing rate, but also interferes with the activities of surrounding neurons. This effect is not limited to the synaptic structure, but exists extensively in a neuron population~\cite{pillow2008spatio}.

To combine the above contributing factors, we update the hidden state of each neuron in a recurrent architecture.
\begin{equation}\label{eq3}
    \mathbf{h}_{i,t} = g(\rm{CONCAT}[\mathbf{h}_{i,t-1} ; \mathbf{s}_{t} ; \mathbf{m}_{i,t-1}])
\end{equation}
where $\mathbf{m}_{i,t}$ denotes the representation of overall message received by neuron $i$ at time $t$. Inspired by the former analysis, we devise a neural network called Neuron Activation Network (NAN) which consists of three jointly trained components to separately model the terms in interest. An overview of network structure is shown in Fig.~\ref{fig:workflow}. We will introduce the details of different functional components in turn.

\subsection{Spatiotemporal Learning of Sensory Environment}
To capture specific patterns in sensory environment, we use convolutional neural network (CNN)~\cite{lecun1998gradient} and thereby acquire an embedding for the stimulus effect. Due to its generality, the usage of CNN can make our model well applied to any structured sensory modalities, like visual and auditory stimulus. As a comparison to the conventional CNN, we add a temporal dimension to the filters considering the temporal correlation of stimulus with neuron spikes. Hence, the sensory environment variable $\mathbf{M}_t$ $\in \mathbb{R}^{W \times H \times \eta}$ denotes a three-dimension matrix, with the width, height and timescale respectively. Notice that $\mathbf{M}_t$ contains history stimulus in the previous $\eta$ time bins. With the above settings, the stimulus factor $\mathbf{s}_{t}$ can be formulated as below:
\begin{align}\label{stim}
\mathbf{s}_{t} = {\rm CNN}(\mathbf{M}_t)
\end{align}

\subsection{Message Passing via Population Graph}
For lack of prior knowledge on neuron connectivity, we construct a fully-connected graph on the neuron population, which means messages can be passed between any two neurons. Each node in the population graph represents a neuron and the edges represent the message between coupling neurons. Our purpose is to learn the message embedding $\mathbf{m}_{i,t}$ of neuron $i$ and thereby update the hidden state vector of each neuron. The embedding representation of edge derives from hidden states of the two associated neurons:
\begin{align}\label{edge2node}
\mathbf{e}_{ij,t}=f_{edge}(\mathbf{h}_{i,t}, \mathbf{h}_{j,t})
\end{align}
where $f_{edge}$ is a neural network that transfers individual neuron representations to edge representations. Each neuron aggregates all the messages it receives from surrounding neurons to update its message embedding $\mathbf{m}_{i,t}$.  Although the population graph is constructed as fully-connected, the relations between neurons are not quite the same, depending on neuron types, and center distances. Therefore, a weighted factor that represents the relation intensity between neurons is needed. Inspired by the Graph Attention  Network (GAT)~\cite{velivckovic2017graph} and gated attention network (GaaN)~\cite{zhang2018gaan},  we come up with an auxiliary neural network that directly learns the relation between neurons with purely the spike train observations. This auxiliary network also adopts the form of message acquisition in Eq.~\ref{edge2node}, but uses a 1-dimension channel output and ultimately gets a weighted matrix $\mathbf{K} \in \mathbb{R}^{N \times N}$, where $\mathbf{K}_{ij}$ denotes neuron $j$'s effect on neuron $i$. To mimic the synaptic integration in a biological neuron~\cite{hao2009arithmetic}, the aggregation of  received messages can be defined as:
\begin{align}\label{eq6}
\mathbf{m}_{i,t}={\rm AGGREGATE}[\mathbf{K}_{ij} f_{node}(\mathbf{e}_{ij,t})]
\end{align}
where $f_{node}$ is a neural network that transfers edge embeddings to node embeddings. Since the spike train observations are dynamic throughout the whole recording time, the learned relation matrix $\mathbf{K}$ can also change by time, which enables our model to support operating on dynamic graphs. This might be useful for the neural recording that spans ultra-long timescales or that consists of multiple periods separated by several time intervals. In summary, the design of neuron relation  learning process makes the whole graph insusceptible to changes in neuron structure and supports online learning.

\subsection{Learning Representation with Mutual Information}
One of our goals is to derive a high quality representation of the neuron hidden state $\mathbf{H}$, which will be beneficial to more downstream applications in neuron coding technology. While a single neuron may be unreliable due to high trial-to-trial variability, the whole neuron population always exhibits a relatively stable neural activity. That means, with a fixed stimulus input, the visual cortex can always encode the stimulus precisely and invariantly. It is intuitively because noises are averaged over the single neurons in the population.  Theoretical studies also have shown that most types of noise are indeed harmless at the population level \cite{abbott1999effect,ecker2011effect,sompolinsky2001population}. So on analysis of single-trial neuron spikes, apart from the exactness of our generated, artificial spikes of single neurons, we try to extract information under this stability across neuron population. 

Based on the aforementioned observations, an underlying working pattern of neuron coding can be constructed. To this end, we assume that with given stimulus, neurons in a population work as a group, so they share information within the neighbourhood to guarantee the most remarkable pattern of the stimulus is captured even considering its own variability. It can be deemed as another implicit way they interfere with each other. These more telling information is preserved in the neuron's hidden state, which is a high dimension space providing intrinsic cerebral condition.

To concretize the measurement, we are inspired to use mutual information for learning representation of high quality as well as eliminating the single neuron uncertainty. Statistically, mutual information can be regarded as the Kullback-Leibler (KL) divergence from the joint density $p(\mathbf{x},\mathbf{y})$ to the product of the marginals $p(\mathbf{x})p(\mathbf{y})$, and thus can be regarded as a measure of statistical dependency between random variables $\mathbf{X}$ and $\mathbf{Y}$. The mutual information can be formally expressed as below:
\begin{equation}
    I(\mathbf{X};\mathbf{Y}) = \iint p(x,y)\log\frac{p(x,y)}{p(x)p(y)}
\end{equation}

As the neurons collaborate with each other, they will preserve a certain kind of similarity both from local neighbours and the entire population. And at the same breath, the representation of the whole graph will be more telling since it is enforced to grasp all of the similarities among neurons. In this paper, we are motivated to maximize the pair-wise neuron mutual information. Theoretical proof can be given to show that this objective can also maximize the mutual information between whole graph and single neurons. We consider a 2-neuron circumstance. The distribution of features can be expressed as $p_1$ and $p_2$ for the two nodes, and that of the whole graph is denoted as $q$. Since $I(p1;q)+I(p2;q) = H(p_1) + H(p_2)-H(p_1|q)-H(p_2|q) \le H(p_1) + H(p_2) - H(p_1p_2|q)$ and $I(p_1;p_2) = H(p_1) + H(p_2) - H(p_1p_2) \le H(p_1) + H(p_2) - H(p_1p_2|q)$, where$H(\cdot)$ is the information entropy function. We can see the two paradigms have identical objective functions. 

Hence, our overall optimization function can be formally expressed as below. It is a combination of the log likelihood of neural sequences with regularization terms:
\begin{equation}
    \mathcal{L} = \log p(\mathbf{Y}|\mathbf{X}) + || \mathbf{K}||+ \sum_{i,j}I(\mathbf{h}_{i,t};\mathbf{h}_{j,t}) 
\end{equation}
where $||\mathbf{K}||$ represents the Euclidean norm of $\mathbf{K}$. Intuitively, the introduction of mutual information maximization can be seen as another regularization other than $||\mathbf{K}||$. While the latter is inspired by the synaptic sparsity, maximizing the mutual information can reduce the redundancy of the high dimension hidden state. Since the spike series is a low dimension display of the hidden state, such a regularization is necessary in acquring high-quality representations.

\section{Experiments}
\begin{figure*}[tb!]
	\centering
    
    \begin{minipage}{0.44\textwidth}
            \subfloat[Reciprocal of MAE]{\includegraphics[width=0.4\linewidth]{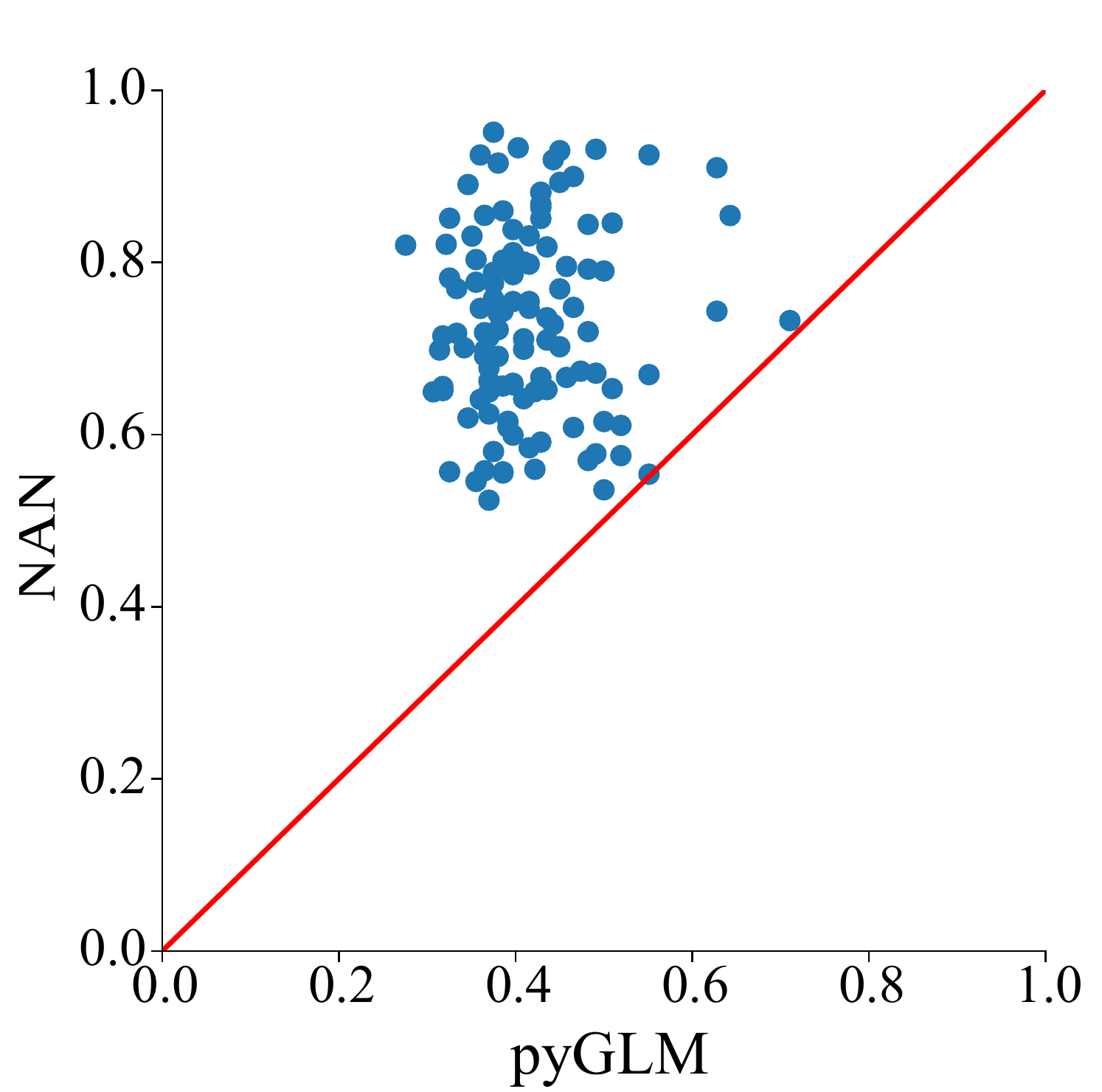}
            \label{fig:mae}}
            \quad
            \subfloat[Fraction of EV]{\includegraphics[width=0.5\linewidth]{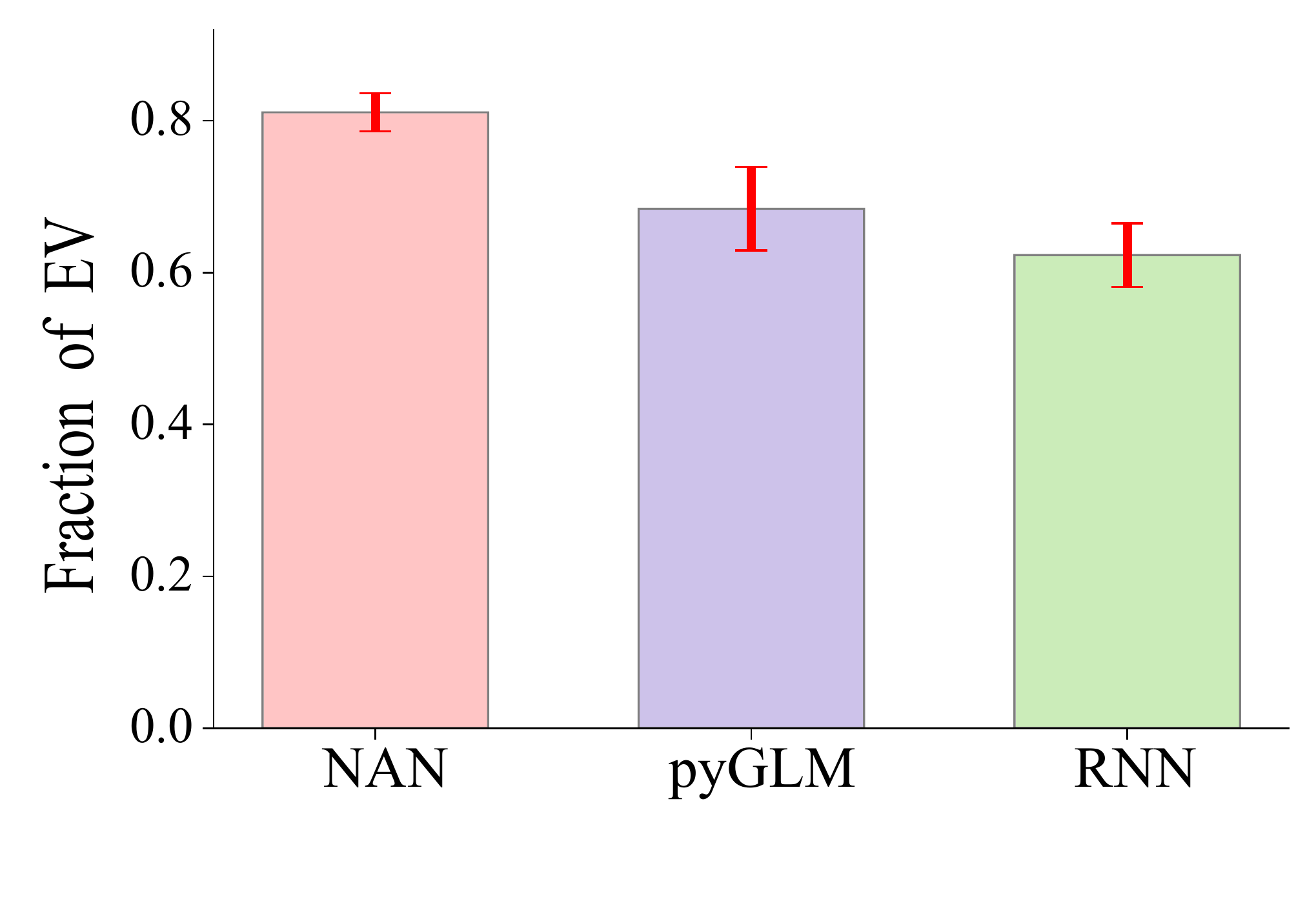}
            \label{fig:expvar}}\\
            \subfloat[Xcorr. on Neuron 18]{\includegraphics[width=0.45\linewidth]{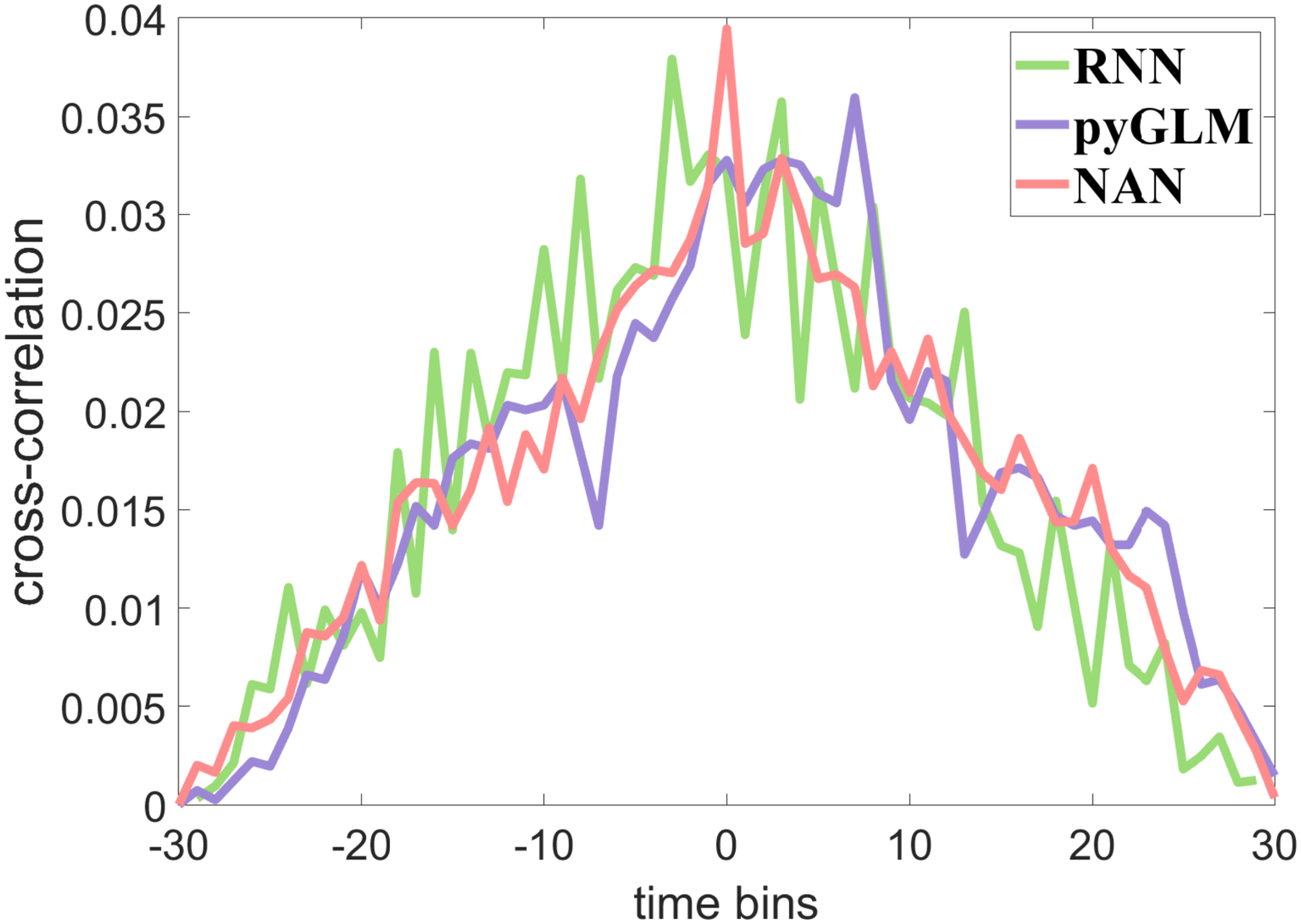}
            \label{fig:neur18xcorr}}\quad
            \subfloat[Xcorr. on Neuron 20]{\includegraphics[width=0.45\linewidth]{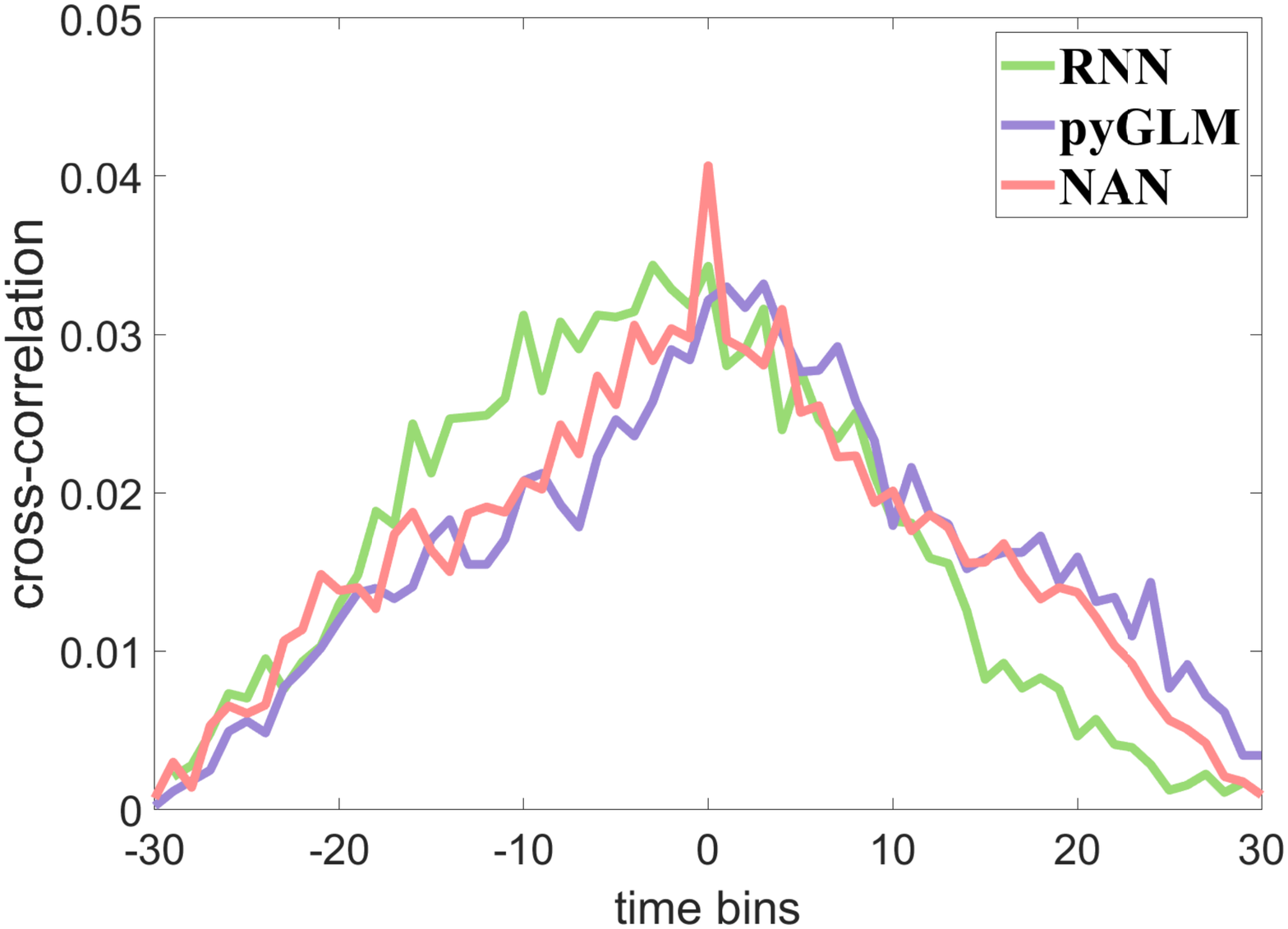}
            \label{fig:neur20xcorr}}
    \end{minipage}
    \begin{minipage}{0.55\textwidth}
        \subfloat[Real (RGC) and simulated (NAN, pyGLM) neuron spike sequences]{\includegraphics[width=1.0\linewidth]{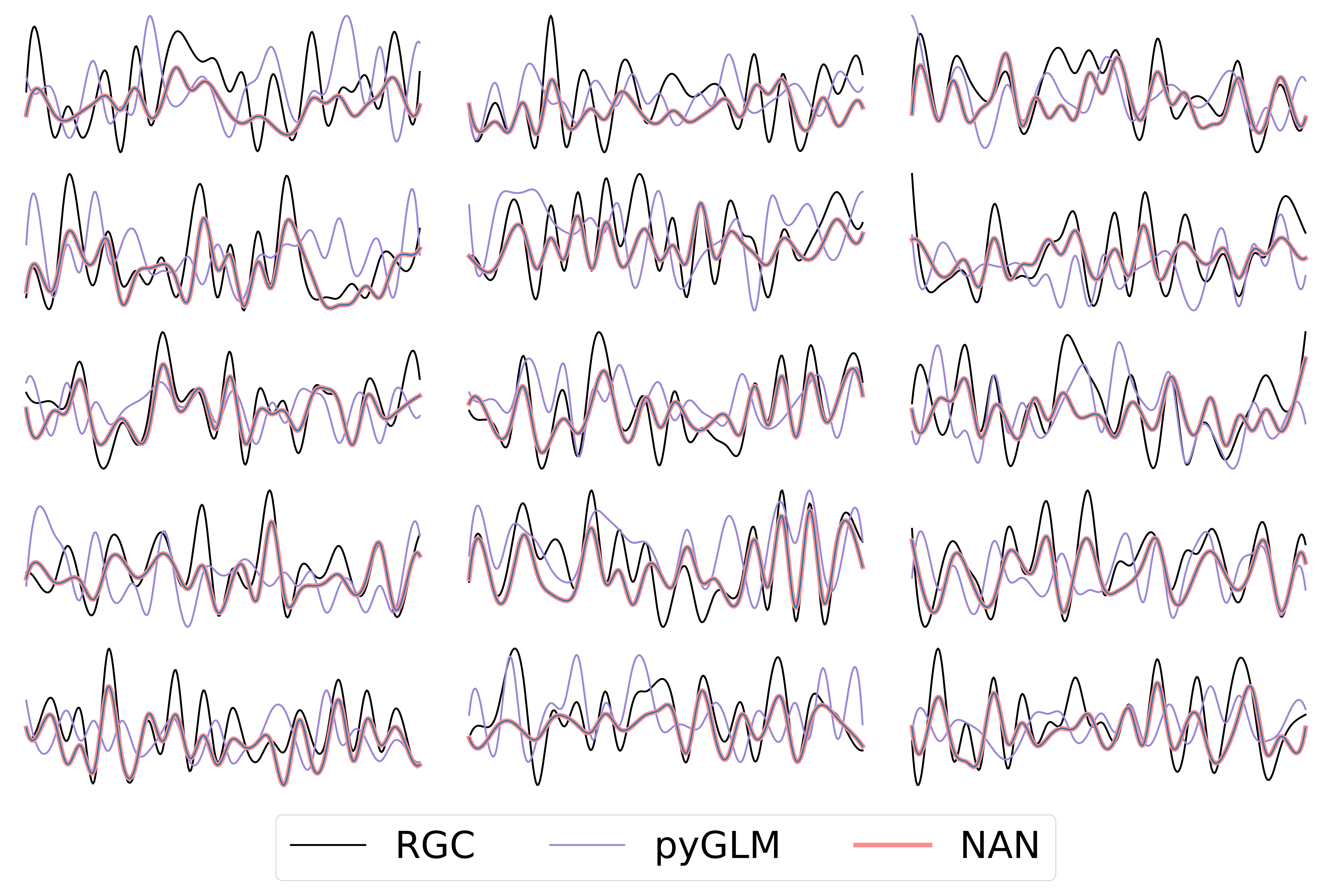}
        \label{fig:responses}}
    \end{minipage}
    \caption{Illustration of model performance. Fig.~1(a) shows the reciprocal of mean absolute error (MAE) comparing NAN and pyGLM. Fig.~1(b) shows the fraction of explainable variance of NAN, GLM and RNN. Fig.~1(c) shows the cross-correlation of the recorded and generated spike trains on Neuron 18 by NAN (red), pyGLM (purple), RNN (green). Fig.~1(d) shows a similarly conducted cross-correlation on Neuron 20. Fig.~1(e) shows the responses of different neurons generated by NAN and pyGLM, both compared with the real retinal ganglion cell spikes. The curves are smoothed with interpolation.}
\end{figure*}

\subsection{Dataset and Protocol}
We fit our models with retinal ganglion cells (RGCs) recorded \textit{in virto} from a segment of isolated macaque monkey retina by multiple electrodes. Based on light response properties and density, these neurons were identified as ON and OFF cells. In the neuron population in study, there are 27 neurons in all, with OFF cells indexed from 0 to 15 and ON cells indexed from 16 to 26 respectively. The retina was stimulated with a spatiotemporal pseudo-random binary sequence refreshing at 120Hz. The spike trains are simultaneously recorded with every stimulus frame which means the neuron population spike train is also sampled at a frequency of 120Hz. This dataset is the same  one as used in~\cite{pillow2008spatio,linderman2016bayesian}. In this paper, the neural recordings are transformed appropriately to spike trains that approximate latent firing rates.

In practice, the first-hand spike data appear in a form of event sequence, with the exact spike generation time recorded.  For simplicity of analysis and data processing, we convert the inconsistent event sequences to simultaneous spike trains on neuron population. We also aggregate the successive time bins at a certain stride to get spike count data. For the stimulus, we concatenate multiple previous stimulus frames on the third dimension, forming a stimulus movie for every time bin in study. All the dataset is partitioned into training, validation and testing set by 10:1:1.

\begin{table*}[t]
    \centering
    \caption{Comparison of cross correlation on the whole neuron population. Each element here denotes the peak of the cross correlation from the generated neural spike series. The first number is its peak value in cross-correlation (in a unit of $10^{-1}$). The integer in brackets is the offset by which the cross correlation deviates from its original position to meet the maximum.}
    \begin{tabular}{cccccccccc}
    \toprule
    models & \# 1&\# 2& \# 3&\# 4&\# 5&\# 6&\# 7&\# 8&\# 9 \\
    \midrule
    pyGLM &0.328(-1)&0.319(-2)&0.337(0)&0.317(-1)&0.316(0)&0.339(0)&0.327(0)&0.331(0)&0.366(0)\\
    RNN&0.337(1)&0.331(2)&0.330(0)&0.334(0)&0.340(6)&0.337(2)&	0.325(0)&0.347(1)&\textbf{0.388(1)}\\
    NAN&\textbf{0.353(0)}&\textbf{0.354(0)}&\textbf{0.349(0)}&\textbf{0.369(0)}&\textbf{0.346(0)}&\textbf{0.353(0)}&\textbf{0.363(0)}&\textbf{0.375(0)}&0.382(0)\\
    \bottomrule
    \end{tabular}
    \begin{tabular}{cccccccccc}
    \toprule
    models & \#10&  \# 11&  \# 12& \# 13& \# 14& \#15& \# 16& \# 17& \# 18\\
    \midrule
    pyGLM &0.321(0)&0.349(0)&0.323(0)&0.325(-1)&0.334(0)&0.323(0)&0.329(-1)&0.343(1)&0.360(7)\\
    RNN&0.334(1)&0.342(1)&0.339(1)&0.342(1)&0.350(0)&0.345(1)&0.342(0)&0.338(0)&0.368(-3)\\
    NAN&\textbf{0.347(0)}&\textbf{0.386(0)}&\textbf{0.364(0)}&\textbf{0.367(0)}&\textbf{0.351(0)}&\textbf{0.361(0)}&\textbf{0.343(0)}&\textbf{0.360(0)}&\textbf{0.395(0)}\\
    \bottomrule
    \end{tabular}
    \begin{tabular}{cccccccccc}
    \toprule
    models & \# 19 & \#20&  \# 21&  \# 22& \# 23& \# 24& \#25& \# 26& \# 27\\
    \midrule
    pyGLM &0.340(1)&0.332(3)&0.330(0)&0.348(4)&0.322(0)&0.338(0)&0.345(1)&0.333(0)&0.337(1)\\
    RNN&0.341(3)&0.344(-3)&0.347(0)&0.343(-1)&0.340(0)&0.349(-2)&\textbf{0.539(3)}&0.334(-1)&0.333(-2)\\
    NAN&\textbf{0.367(0)}&\textbf{0.407(0)}&\textbf{0.396(0)}&\textbf{0.368(0)}&\textbf{0.371(0)}&\textbf{0.359(0)}&0.366(0)&\textbf{0.372(0)}&\textbf{0.359(0)}\\
    \bottomrule
    \end{tabular}
    
    \label{table1}
\end{table*}

\subsection{Model Setup}
We clarify the architecture and parameters used in our proposed method, Neuron Activation Network (NAN).  This model has  an input of  spike history $\textbf{X} \in \mathbb{R}^{N\times T}$. Here 
we set $N=27$ and $T=20$. The first component of the NAN is a neural network that captures the relations between single neuron units, a.k.a inferring the relation matrix $\textbf{K} \in \mathbb{R}^{N \times N}$. The node-edge translation function $\emph{f}_{edge}$ takes the concatenation of the spike sequences (with a length of 20 time slots) of the two vertices of an edge as input. The architecture of  $\emph{f}_{edge}$ is a fully connected neural network with input dimension of 60 and has one hidden layer with dimension of  128 and an output layer with dimension of 1. Hence the relativity of every two neurons is demonstrated by a scalar. In the spatiotemporal learning process,  a 3D Convnet is proposed, which takes a tensor $M \in \mathbb{R}^{W \times H \times \eta}$ as input. Here $W = H = 5$  which means our input is of size 5 $\times$ 5. And  $\eta$ is selected as 20.  The convolution kernel  is of size $3 \times 3 \times 20$. The output of the network is vectorized as the representation of stimulus's impact on the neurons.  The message aggregation process involves an edge-node translation function $\emph{f}_{node}$ and its counterpart $\emph{f}_{edge}$.  Similar as the one stated above,  $\emph{f}_{edge}$ here only has a mere modification that the output dimension is 128, deemed as the representation of edge.  $\emph{f}_{node}$ takes this representation in and uses a two layer ELU with output dimension 128. Its output is multiplied by the corresponding element in $\textbf{K}$ (which acts as attention mechanism) according to Eq.~\ref{eq6}, the output of which is our  learned representation of the impact on a neuron by its neighbours. So with the stimulus representation and the surrounding representation, we can refer to Eq.~\ref{eq3} and update the hidden state representation of each neuron. Function $g$ used in Eq.~\ref{eq3} is an MLP mapping the concatenation of $\mathbf{h}_{i,t-1}$,$ \mathbf{s}_{t}$ and $ \mathbf{m}_{i,t-1}$ to a 256 ($n_h$)-dimension vector.  We train the model on Nvidia GeForce RTX 2080Ti with 10GB memory. Although the given equations are in neuron granularity, our implementation takes advantage of matrix computation and can be parallelized on a whole population scale, which contributes to the scalability of the model.

In the optimization objective we adopt, mutual information is notoriously difficult to compute, particularly in continuous
and high-dimensional settings~\cite{hjelm2018learning}. We leverage a similar design described in previous works ~\cite{velivckovic2018deep}. In actuality, any divergence between the joint and the product of marginals should be sufficient for estimating and maximization mutual information. Hence we adopt a standard binary cross-entropy loss between positive and negative samples. Considering the adversarial learning prerequisite, we divide the samples into positive and negative ones according to the relation intensity between two neurons. The neuron pair with higher relation intensity indicates they share more information with each other. Thus we define the variant mutual information loss as below,
\begin{equation}
    {\rm MILoss}= \log \mathcal{D}(\mathbf{h}_{i,t};\mathbf{h}_{j,t}) + \log (1 - \mathcal{D}(\mathbf{h}_{i,t};\mathbf{h}_{k,t}))
\end{equation}
where $\mathbf{h}_{j,t}$ and $\mathbf{h}_{k,t}$ are positive and negative samples respectively for neuron $i$. $\mathcal{D}$ is a discriminative network with a scalar output. To guarantee the symmetry of mutual information ($I(x;y) = I(y;x)$), network $\mathcal{D}$ is a bilinear regressor, which mathematically conforms with the commutative law.

\subsection{Model Performance}
With our Neuron Activation Network (NAN), we can predict a spike sequence given spike history and the stimulus that elicits the spikes. We train a Bayesian inference based model pyGLM~\cite{linderman2016bayesian} to show the superiority of NAN. We calculate the mean absolute error (MAE) of the predicted spike sequence with the targeted one. The MAE is averaged across all the 27 neurons in different time bins. Fig.~\ref{fig:mae} compares the reciprocal of MAE with NAN and pyGLM. It suggests that our NAN consistently outperforms pyGLM in almost every single time bin. 

To further quantitatively evaluate the accuracy of our predicted spike sequences, we calculate the fraction of explainable variance (EV), which has also been used in previous literature~\cite{heitman2016testing,batty2016multilayer}. This metric compares the generated spike sequence with the recorded one by using simulated firing rate as a predictor of the recorded rate. The fraction of explainable variance is computed according to the following definition:
\begin{align}
    F(r,r_s)=1-\frac{\sum_t(r(t)-r_s(t))^2}{\sum_t(r(t)-\mu)^2}
\end{align}
where $r(t)$ is the recorded spike count data, $r_s(t)$ is the generated spike sequence, and $\mu$ is the average recorded rate. 
Fig.~\ref{fig:expvar} shows the fraction of explainable variance. We compare our NAN with pyGLM and an RNN baseline. The choice of baseline is validated in \cite{batty2016multilayer}, as varying RNN architectures lead to similar levels of performance. We find that our model captures around 80\% of the explainable variance and gives numerical evidence in our improvement for high-quality spike train generation. 

\begin{figure}[tb!]
  \centering
  \includegraphics[width=0.5\textwidth]{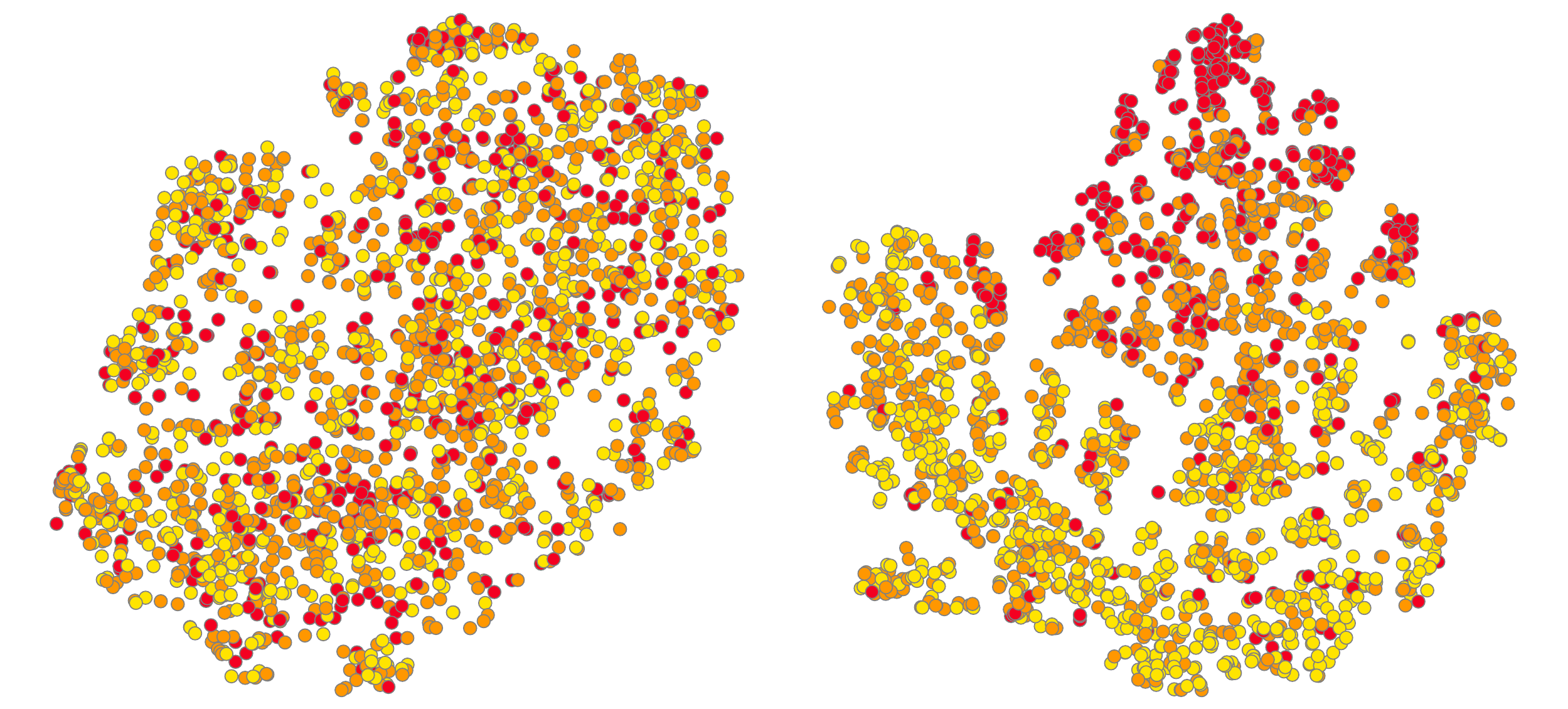}
   \vspace{-10pt}
  \caption{Juxtaposed t-SNE embeddings of the neuron features from a randomly initialized NAN ($\mathbf{left}$) and a trained NAN model ($\mathbf{right}$). These points contain all the neurons across large scale of timeline. Color of each node represents neuron activity (high: red / medium: orange / low: yellow) of the neuron based on its spiking rate observation. With NAN, neurons of different activation conditions are  clustered.}
  \label{fig:hidden}
\end{figure}

MAE and fraction of EV are static metrics focusing on single time bin and single neuron respectively. Actually, NAN is more powerful as a spike dynamics generator. From the generated spike sequences shown in Fig. \ref{fig:responses}, we can have a global view of the temporal 
performance of NAN. NAN produces a spike curve that is geometrically correlated with the observed neuron recordings, showing its capability beyond only giving a numerical approximation of the firing rates within a single time bin. As a single-trial spike generation is always varied due to fluctuation in individual neurons, to predict the trend of a spiking activity is much more an important and illuminative work. Moreover, superior to the pyGLM model depicted in Fig.~\ref{fig:responses}, our  model has the ability to make long-term predictions accurately without an obvious depravity in performance or displacement of spike curve as shown in pyGLM. 

This displacement of spike curve can be modeled with a statistical method called cross correlation which is a standard metric in estimating the degree to which two series are correlated and has been used extensively in neuron spike analysis. The complete cross correlation of generated neural spikes are shown in Table.~\ref{table1}.  For display purpose, we take Neuron 18 and Neuron 20 for example. Fig.~\ref{fig:neur18xcorr} and Fig.~\ref{fig:neur20xcorr} show the cross-correlation depending on time bin lags. First, we subtract the time-varying mean firing rate of two spikes to eliminate correlations induced by similarity in mean firing rates. Then we calculate the time-lagged cross-correlations for the spike trains generated by NAN, pyGLM and RNN respectively with the recorded spike series. In the figures, our proposed Neural Activation Network consistently achieve the highest correlation with target spike train on zero time lag both within its own spike series and compared with other two simulated spike series. Meanwhile, as shown in Table.~\ref{table1}, both pyGLM and RNN always suffer from a correlation peak with a certain time lag offset, which means the simulated spike series by these models tend to show displacement from the target spike series and entail extra effort to be adjusted.  


\addtocounter{figure}{-1}
Moreover, the feature vector extracted by NAN is also of high quality. Fig.~\ref{fig:hidden} shows the initial and learned embeddings for all the neurons in the population sampled from different time bins. The nodes are classified into different colors according to its activation condition (firing rate). For the initial states, neurons with various firing rates are mingled with each other. After extracting features using NAN, nodes are clearly clustered. For instance, the neurons with high firing rates (denoted by red nodes) concentrate in the top of the figure. From the embeddings we have learned, we also observe that the neuron activation process is continuously distributed. The transition of neuron activation condition is featured in our embeddings, as displayed by the gradual change in color from peripheral to central.

\addtocounter{figure}{1}
\begin{figure}[tb!]
  \centering
  \includegraphics[width=0.32\textwidth]{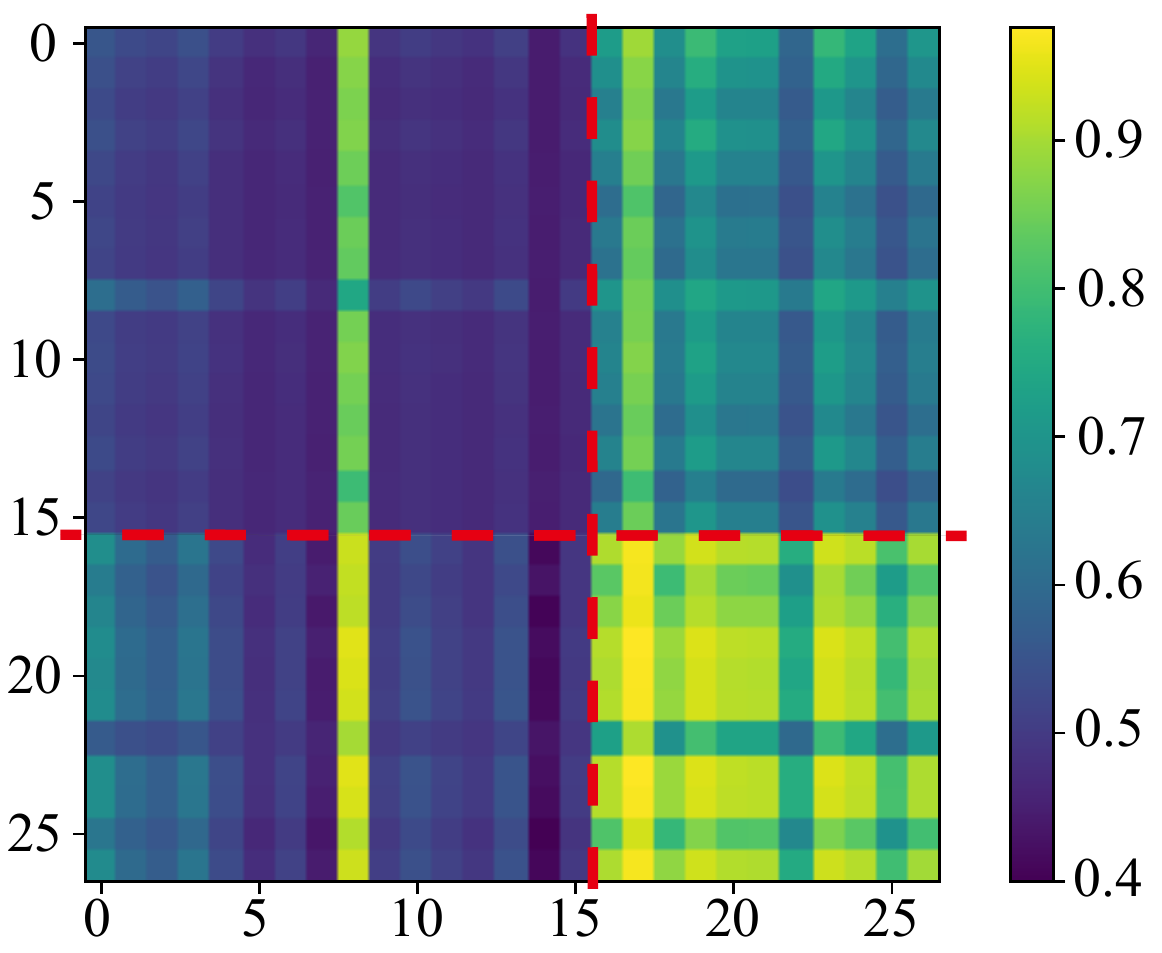}
  \caption{Neuron relation intensity matrix. A brighter color signifies a stronger relation between neurons. Red dashed lines separate neurons of different types.}
  \label{fig:edge_vis}
\end{figure}
\begin{figure}[tb!]
  \centering
  \includegraphics[width=0.4\textwidth]{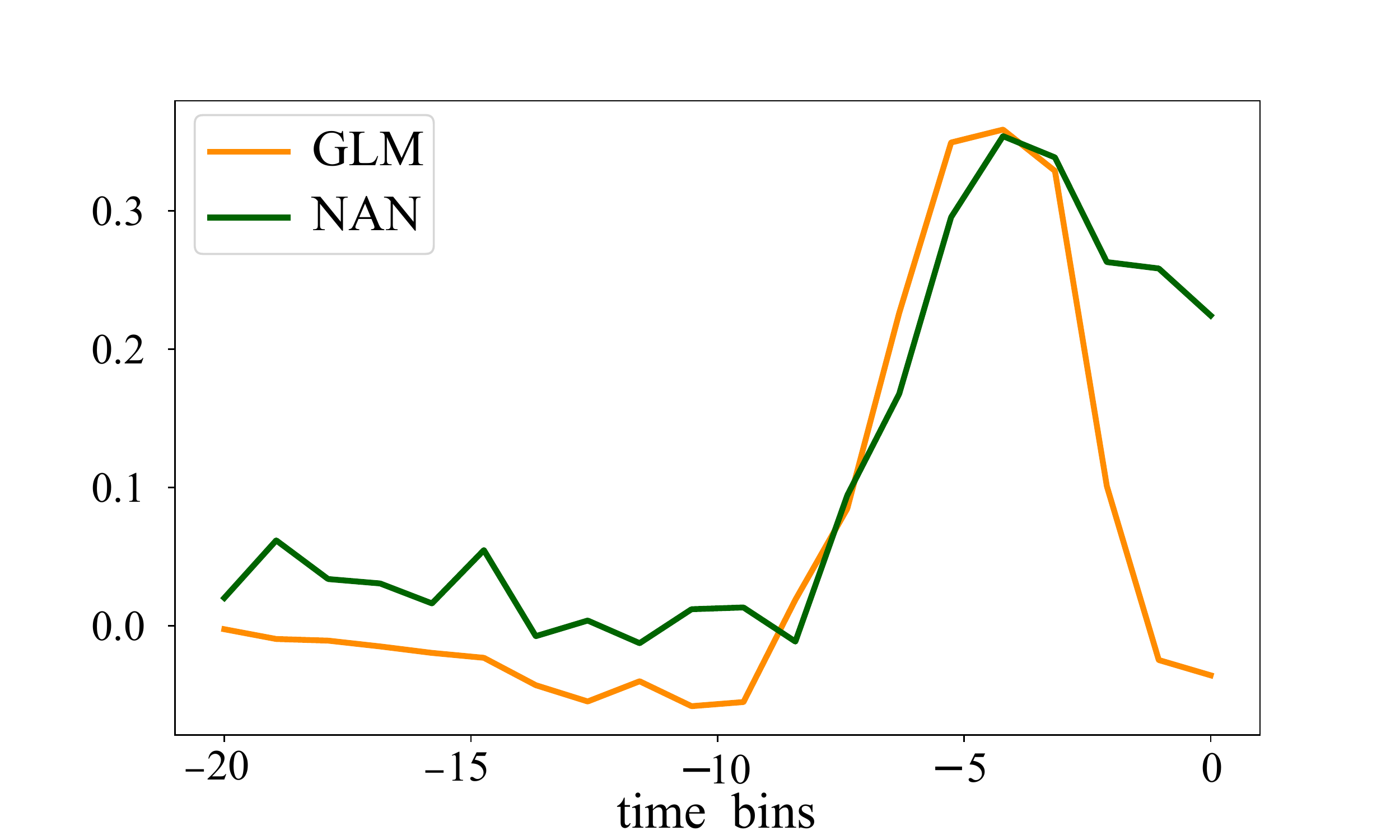}
  \caption{Temporal filter extracted from spatio-temporal filter in sensory network. Our filter is displayed in green  compared with that derived from GLM in orange.}
  \label{fig:temporal}
\end{figure}

\subsection{Interpretable Insights}
Due to the black-box characteristic of deep neural networks, it is a challenging task for one to gain insight into the way our model learns from neural spike data. However, it is a most important part in neuroscience — we not only expect the model itself to learn how neuron population works, but also want to extract and disclose knowledge explicitly and facilitate human-level comprehension. On this concern, our model has made its first strides in explaining the working mechanism of spike  generation. 

\textbf{Learning Neuron Relation}
In virtue of the design of neuron relation learning process, we are able to dynamically infer the neuron relation intensity matrix $\mathbf{K}$. In Fig.~\ref{fig:edge_vis}, we visualize the matrix at a specific time bin. A similar plot on the same dataset can be found in ~\cite{linderman2016bayesian}. The brightness indicates the relation intensity between neuron pairs. We find that the whole matrix is segmented into four orthogons according to the difference in colors, which is made more remarkable by manually putting two red dashed lines on the border of each orthogon. Notice that as described above, the 0-15 cells are classified as OFF cells and 16-26 cells are ON cells. The borders in the matrix exactly split the two neuron types. It suggests that our model can learn the type of cells as well as the relations between different types with no prior and in an unsupervised fashion merely based on the pattern in spike observations.

\textbf{Learning Temporal Filter}
In this paper, we apply a spatiotemporal filtering to the input sensory environment. With this implementation, individual neurons are enabled to capture both spatial and temporal features from stimulus. Fig.~\ref{fig:temporal} shows the temporal dimension of the spatiotemporal filter in the 3D ConvNet. More specifically, we apply a rank-one factorization for the filter parameter matrix $\textbf{Z}^{w \times h \times \eta}$ via Canonical Polyadic (CP) Decomposition. We compare the inferred temporal dimension of the spatiotemporal filter to that derived from GLM~\cite{pillow2008spatio} and find that the geometrical characteristics are strongly correlated between the two derived filter curves. It means our model has a credible property in discovering temporal relations from stimulus and ensures that our model has decoded the stimulus in a right and meaningful way. This approximation also provides us another option to infer the temporal effects on neuron spike generation beyond GLM.

\subsection{Discussion}
The experiments show that NAN is not only replacing a black-box model (cerebral information processing system) with another (artificial neural network), but also to integrate human designs and insights into a potent non-linear system. Our method is prominent in generating spike sequences with high fidelity as well as disclosing latent characteristics under neural circuits. The former ensures our model has captured the spike generation mechanism on the whole. The latter, which is more expressive and exciting, suggests that we are able to extract neural information from the model. This strategy of constructing a deep network to capture neural mechanism function and then constraining it to better understand the mechanism will be useful in a variety of neural systems in response to much more complex stimuli. 

Nevertheless, it is still an onerous task to further explore and exploit more valuable information in spike train data. First, we are still not able to uncover anatomical connections between cells. Although we have revealed characteristics of individual neurons and explain the relation intensity to some degree, we cannot tell if the inferred intensity actually reflects a connection between neurons or just depending on shared input noise. In fact, decoding neuron relations is challenging for lack of ground-truth data in practice, which makes the problem a seemingly unsolvable one. It needs more than the spike train observations and a more specialized neuro-probing technique that might contribute to our knowledge of neuron population. Second, being consistent with previous traditional works, our proposed method also treats the spike train data as multidimensional time series, where a core problem is to determine the granularity by which we aggregate discrete time bins. This will unavoidably lose several mechanisms known to exist in neurons (for example refractory period). One possible solution is to treat the spike train data as event sequences. A flourishing technique called event sequence analysis is helpful in this scenario. Techinques combining the event sequences with generative models like variational auto-encoder or generative adversarial network e.g. WGANTPP~\cite{xiao2017wasserstein}, is a promising direction, which we leave for future work.

\section{Conclusion}
We have tried to answer a fundamental and intriguing problem in cognitive neuroscience -- how do spike trains form in a neuron population and how can we disclose underlying neuron characteristics using spike train observations? We find that traditional methods are usually plagued by a lack of flexibility or scalability. Few researches have turned to adopt a neural network approach, which has a stronger ability for nonlinear fitting. We believe that using neural networks, we can get easier access to the spike eliciting generation mechanism in neuron populations. 

To this end, we propose the Neuron Activation Network (NAN) to model the contributing factors and emulate spiking activities. Experiments have shown that our model can generate high quality spike sequences as well as learn representations for neuron hidden state. In summary, this paper makes a progress in combining perception in neuron spike generation mechanism with neural network models to exploit more valuable knowledge in cognitive neuroscience. We hope this model can give light to and co-evolve with development in neuroscience.


\bibliographystyle{plain}
\bibliography{survey}



%








\end{document}